\documentclass[11pt,a4paper]{article}

\usepackage[utf8]{inputenc}
\usepackage[T1]{fontenc}
\usepackage{amsmath,amssymb,amsthm}
\usepackage{graphicx}
\usepackage{hyperref}
\usepackage{authblk}
\usepackage{setspace}
\usepackage{geometry}
\usepackage{titlesec}
\usepackage{indentfirst}

\renewcommand{\thesection}{\Roman{section}}
\titleformat{\section}
  {\centering\normalfont\bfseries}
  {\thesection.}
  {0.8em}
  {\MakeUppercase}
  
\title{\vspace{-1.2cm}
\textbf{Quasinormal modes of scalar and Maxwell field perturbations coupled to the Einstein tensor in generalized Nariai spacetimes}
}

\author[1,2]{Xin Qin\thanks{Corresponding author: qx@hnust.edu.cn}}
\author[3]{Xiongying Qiao}
\author[4,5]{Qiyuan Pan}
\author[1,2]{Xiangyun Fu}
\author[1,2]{Xuan Zhou}
\author[1,2]{Ying Yang}

\affil[1]{School of Physics and Electronic Science, Hunan University of Science and Technology, Xiangtan 411201, China}
\affil[2]{Key Laboratory of Intelligent Sensors and Advanced Sensing Materials of Hunan Province, Hunan University of Science and Technology, Xiangtan 411201, China}
\affil[3]{Department of Physics, Xinzhou Normal University, Xinzhou, Shanxi 034000, China}
\affil[4]{Department of Physics, Institute of Interdisciplinary Studies, Key Laboratory of Low Dimensional Quantum Structures and Quantum Control of Ministry of Education, Synergetic Innovation Center for Quantum Effects and Applications, and Hunan Research Center of the Basic Discipline for Quantum Effects and Technologies, Hunan Normal University, Changsha, Hunan 410081, China}
\affil[5]{Center for Gravitation and Cosmology, College of Physical Science and Technology, Yangzhou University, Yangzhou 225009, China}

\date{}

\begin{document}

\maketitle
\vspace{-0.25cm}

\begin{center}
\textbf{Abstract}
\end{center}

\vspace{-0.15cm}

\begin{center}
\begin{minipage}{0.86\textwidth}
\begin{spacing}{1.25}
\noindent
We investigate the quasinormal modes of scalar and Maxwell field perturbations coupled to the Einstein tensor in generalized Nariai spacetimes. Our results show that, for both types of perturbations, the coupling introduces different critical values, which separate the frequency spectrum into distinct branches. Near these critical values, the square-root term that determines $\omega_R^2$ may change sign, giving rise to a parameter interval in which the modes are purely imaginary. Away from this regime, the coupling affects the oscillatory parts of the two fields in opposite ways: $\omega_R^2$ generally increases with the coupling constant $\eta$ for the scalar field, whereas it decreases with $\eta$ for the Maxwell field. The magnetic charge tends to enhance the oscillatory response, while increasing the spacetime dimension narrows the purely imaginary regime. This comparison shows analytically that the same curvature coupling can affect scalar and Maxwell perturbations in qualitatively different ways.
\end{spacing}
\end{minipage}
\end{center}

\clearpage

\section{Introduction}

The detection of gravitational waves from compact binary coalescences by LIGO has made the ringdown phase of black holes an important observational window onto strong field gravity \cite{LG1,LG2,LG3}. This stage carries information about the mass and spin of the remnant black hole, as well as possible deviations from the predictions of general relativity. In realistic astrophysical environments, black holes are not isolated systems, but are affected by surrounding matter and external perturbations. Within linear perturbation theory, the corresponding response is characterized by quasinormal modes (QNMs), whose frequencies are discrete and complex: the real part gives the oscillation frequency, while the imaginary part determines the damping or growth rate. These frequencies depend on the parameters of the background black hole, the type of perturbing field, the imposed boundary conditions, and possible nonminimal couplings. Recent studies have extended QNM analyses to various modified gravity models and nonstandard black hole backgrounds \cite{QM1,QM2,QM3,QM4,QM5,QM6,QM7,QM8,QM9,QM10,QM11,QM12}. Thus, QNM frequencies provide a useful probe of both the background geometry and possible curvature couplings of perturbing fields.

Higher dimensional spacetimes arise naturally in string theory, brane world scenarios, and the AdS/CFT correspondence \cite{xll,msj1,msj2,msj3,ac1,ac2}. These motivations have led to extensive studies of higher dimensional black holes and their QNMs. Among the known exact solutions, Nariai spacetimes and their generalizations are useful because their product structure allows the perturbation equations to be reduced to tractable radial problems. The four dimensional Nariai solution has the geometry $dS_2\times S^2$, namely the direct product of two constant curvature spaces, and can be regarded as the near horizon geometry of the Schwarzschild-de Sitter black hole in the limit where the black-hole and cosmological horizons coincide \cite{Nariai1}. Charged higher dimensional Nariai spacetimes were later constructed \cite{Nariai2}, with geometry $dS_2\times{S^{n-2}}$. More general Nariai solutions were subsequently obtained as products of $dS_2$ with several two spheres \cite{Nariai3}, $dS_2\times{S^2}\times\cdots\times{S^2}$. These solutions can carry not only electric charge but also several magnetic charges, and therefore possess a richer geometric and electromagnetic structure than the usual higher dimensional Nariai backgrounds. Because the perturbation equations are analytically tractable in this background, generalized Nariai spacetimes provide a useful setting for examining how nonminimal curvature couplings affect QNM frequencies.

QNMs are determined not only by the background geometry, but also by possible nonminimal couplings between the perturbing field and the curvature. A representative example is the coupling of a scalar field to the Einstein tensor \cite{oh1,oh2,oh3,oh4,oh5,oh6}. This coupling is of particular interest because it introduces curvature dependent interactions while keeping the equations of motion second order, thereby avoiding the unstable degrees of freedom that may arise in generic higher derivative theories. In black hole backgrounds, it modifies the effective geometry and the effective potential experienced by the scalar field, thereby changing the QNM frequencies and stability properties. Previous studies in Reissner-Nordström, Reissner-Nordström-AdS, and asymptotically de Sitter black hole spacetimes have shown that the nonminimal coupling between a scalar field and the Einstein tensor can substantially change the perturbation dynamics \cite{oh7,oh8,oh9}. In particular, for massive scalar fields or in the presence of a positive cosmological constant, unstable regions may appear, controlled jointly by the background parameters, the field mass, and the coupling strength. These results show that Einstein tensor coupling can affect scalar
perturbations in a nontrivial way, and motivate the corresponding study for vector fields.

Nonminimal curvature couplings of vector fields have been studied in several contexts, including electromagnetic fields coupled to the Weyl tensor, vector tensor theories, and Horndeski vector tensor gravity \cite{oh10,oh11,oh12}. These studies show that such couplings can modify the effective potential, QNM frequencies, propagation properties, and stability of vector perturbations in black hole backgrounds. Moreover, in a four dimensional Reissner-Nordström black hole background, the perturbation equation for a vector field coupled to the Einstein tensor depends not only on the coupling parameter, but also sensitively on the parity of the perturbation \cite{oh13}. These results indicate that curvature couplings can alter the standard propagation properties of vector perturbations, motivating the study of the Einstein tensor coupling in less explored higher dimensional backgrounds. In this paper, scalar and Maxwell field perturbations coupled to the Einstein tensor are investigated in higher dimensional charged generalized Nariai spacetimes. We focus on how the coupling parameter, background electromagnetic parameters, and spacetime dimension affect the QNM frequencies and the stability of the perturbations.

The structure of this paper is as follows. In Sec. 2, we briefly introduce the higher dimensional generalization of the charged Nariai spacetime. In Sec. 3, the QNM frequencies of scalar field perturbations coupled to the Einstein tensor are derived, and the effects of the coupling on perturbation stability are examined.  In Sec. 4, the QNM frequencies of Maxwell field perturbations coupled to the Einstein tensor are obtained on the same background, and the influence of the coupling on the corresponding spectra is analyzed. Finally, our conclusions are given in Sec. 5.

\section{Generalized Nariai Spacetimes}
In this section, we briefly review the higher dimensional generalization of the charged Nariai spacetime \cite{Nariai3}. The action for Einstein-Maxwell theory with a cosmological constant $\Lambda$ in $D$ dimensional spacetime is 
\begin{equation}\label{action}
S=\int\sqrt{-g}\left[\mathcal{R}-(D-2)\Lambda-\frac{1}{4}\mathcal{F}^{cd}\mathcal{F}_{cd}\right],
\end{equation}
where $\mathcal{R}$ is the Ricci scalar and $\mathcal{F}_{cd}$ is the background electromagnetic field strength. The background gauge potential is chosen as
\begin{eqnarray}
\bar{A}=Q_1rdt+\sum_{j=2}^{d}Q_jR_j^2\cos\theta_jd\phi_j.
\end{eqnarray}
Here, $Q_1$ denotes the electric charge, while $Q_j$ denote the magnetic charges associated with the two sphere factors. varying the action with respect to the metric and the gauge potential gives the field equations
\begin{eqnarray}
\mathcal{R}_{ab}-\frac{1}{2}\mathcal{F}_a{}^c\mathcal{F}_{bc}=\frac{g_{ab}}{2}\left[\mathcal{R}-(D-2)\Lambda-\frac{1}{4}\mathcal{F}_{cd}\mathcal{F}^{cd}\right],
\end{eqnarray}
and $\nabla^{a}\mathcal{F}_{ad}=0$. A static solution of these equations is given by the direct product of the two dimensional de Sitter spacetime and $(d-1)$ two spheres, with $d=D/2$ \cite{Nariai3}. The metric can be written as
\begin{equation}\label{Nariai}
ds^2=-f(r)dt^2+\frac{1}{f(r)}dr^2+\sum_{j=2}^{d}R_{j}^2\left(d\theta_j^{2}+\sin^2{\theta_j}d\phi_j^2\right),
\end{equation}
where the lapse function is
\begin{equation}
f(r)=1-\frac{r^2}{R_1^2},
\end{equation}
$R_{1}$ and $R_{j}$ are constants related to charges
\begin{eqnarray}\label{the electric and magnetic charges}
&&R_{1}=\left[{\Lambda-\frac{1}{2}Q_{1}^2+\frac{Q}{2(D-2)}}\right]^{-\frac{1}{2}},  \\ \nonumber
&&R_{j}=\left[{\Lambda+\frac{1}{2}Q_{j}^2+\frac{Q}{2(D-2)}}\right]^{-\frac{1}{2}},
\end{eqnarray}
where $Q$ is given by
\begin{equation}\label{Q}
Q\equiv{Q_{1}^2-\sum_{j=2}^dQ_{j}^2}.
\end{equation}
This solution represents a higher dimensional generalization of the charged Nariai spacetime, reducing to the four dimensional charged Nariai solution for $D=4$ \cite{Nariai1}. The hypersurfaces $r=\pm{R_1}$ are the Killing horizons of the two dimensional de Sitter factor. We will consider scalar and Maxwell perturbations coupled to the Einstein tensor on this background and obtain their QNMs. 

The allowed QNMs depend crucially on the boundary conditions imposed at the two Killing horizons \cite{Nariai3}. Condition (I) imposes an outgoing wave at $-R_1$ and an ingoing wave at $r=R_1$, whereas condition (IV) imposes the reverse. Conditions (II) and (III) impose purely outgoing and purely ingoing waves at both horizons, respectively. It will be shown that only conditions (I) and (IV) lead to well defined QNM spectra, as in the uncoupled case.

\section{Quasinormal modes of Scalar Field Perturbations Coupled to the Einstein Tensor}

We now consider scalar field perturbations coupled to the Einstein tensor in the generalized Nariai background. Let $\phi$ be a scalar field with mass $\mu$, whose action is
\begin{equation}\label{Action-scalar}
\begin{split}
S_m &= \frac{1}{16\pi} \int d^Dx\sqrt{-g} \, \Bigl[ -\bigl(g^{\mu\nu} + \eta G^{\mu\nu} \bigr) \partial_\mu \Phi \, \partial_\nu \Phi- 
\mu^2\Phi^2 \Bigr],
\end{split}
\end{equation}
where $\eta$ is the coupling constant. The term $\eta G^{\mu\nu}\partial_\mu \Phi \, \partial_\nu \Phi$ describes a nonminimal derivative coupling of the scalar field to the background Einstein tensor. Varying the action with respect to $\Phi$ gives the modified Klein-Gordon equation
\begin{equation}\label{KGE}
\frac{1}{\sqrt{-g}} \,\partial_{\mu} \left[\sqrt{-g} \left(g^{\mu\nu}+\eta{G^{\mu\nu}}\right) \,\partial_{\nu} \right]\Phi \,=\,\mu^{2}\Phi \,.
\end{equation}
To cast the wave equation into a Schrödinger type radial form, the tortoise coordinate $r_{\star}$ is introduced by
\begin{equation}\label{TC}
dr_{\star}=\frac{1}{f(r)}\,dr. 
\end{equation}
The metric can then be written as
\begin{equation}\label{FF1}
ds^{2}=\frac{1}{\cosh^{2}( r_{\star}/R_{1})}\,\left(-dt^{2} + dr_{\star}^{2}\right) + \sum_{j=2}^{d}R_{j}^{2}\,d\Omega_{j}^{2} \,.
\end{equation}
Substituting the line element into Eq. \eqref{KGE}, we obtain
\begin{equation}\label{KGE1}
\begin{split}
\left\{ \cosh^{2}\left( r_{\star}/R_{1}\right)\left(1-\eta{\sum\limits_{m=2}^{d}\frac{1}{R_{m}^{2}}}\right)  \left(\partial_{r_{\star}}^{2}-\partial_{t}^{2}\right)  -\mu^{2} 
\right. 
\left. +\sum\limits_{j=2}^{d}\frac{h_j(\eta)}{R_{j}^2}\Delta_{j}\right\}\Phi =0 \,,
\end{split}
\end{equation}
where
\begin{flalign}
&\Delta_{j} \equiv \frac{1}{\sin\theta_{j}} \partial_{\theta_{j}}(\sin\theta_{j}\,\partial_{\theta_{j}}) \,+\,\frac{1}{\sin^{2}\theta_{j}} \partial^{2}_{\phi_{j}},  \nonumber \\
&h_j(\eta)=1+\eta{\left(\frac{1}{R_{j}^{2}}-\frac{1}{R_{1}^{2}}-\sum\limits_{k=2}^{d}\frac{1}{R_{k}^{2}}\right)}.
\end{flalign}
We expand the scalar field $\Phi$ as
\begin{equation}\label{ExpansionSacalar}
\Phi=\int d\omega \sum_{\ell, m}\phi^{\omega}_{\ell,m}(r_{\star}) e^{-i\omega t}\,\mathcal{Y}_{\ell,m} ,
\end{equation}
with
\begin{equation}\label{YLM}
  \mathcal{Y}_{\ell,m} = \prod_{j=2}^d \, Y_{\ell_j}^{m_j}(\theta_j,\phi_j)  \,.
\end{equation}
Here $\ell$ and $m$ collectively denote the sets of angular quantum numbers
$\{\ell_j\}_{j=2}^{d}$ and $\{m_j\}_{j=2}^{d}$, respectively. This yields the following ordinary differential equation for the radial
function $\phi^{\omega}_{\ell,m}$:
\begin{equation}\label{WE}
\left[ \frac{d^{2}}{dr_{\star}^{2}}+\omega^{2}-\frac{\sum\limits_{j=2}^d \frac{\ell_{j}(\ell_j + 1)}{R_j^2}h_j(\eta)+\mu^2}{\left(1-\eta{\sum\limits_{m=2}^{d}\frac{1}{R_{m}^{2}}}\right)\cosh^{2}\left( \frac{r_{\star}}{R_{1}}\right)}  \right] \phi^{\omega}_{\ell,m} = 0 \,.
\end{equation}
A dimensionless variable is introduced as
 \begin{align}\label{y-function}
 y=\frac{1}{2}+\frac{1}{2}\tanh(r_{\star}/R_{1}),
  \end{align}
which maps $r_{\star}=+\infty$ to $y=1$ and $r_{\star}=-\infty$ to $y=0$. To reduce the radial equation to a hypergeometric form, the radial component is written as
 \begin{align}\label{defined function-1}
 \phi^{\omega}_{\ell,m}=y^\rho(1-y)^{\sigma}G(y).
 \end{align}
The function $G(y)$ then satisfies
\begin{align}
  y\left(1-y\right)\frac{d^2{G}}{dy^2}+\left[c-y\left(a+b+1\right)\right]\frac{dG}{dy}-abG=0.
\end{align}
For compactness, we define
\begin{equation}
\Xi_s(\eta)=R_1^2\frac{\mu^2+\sum_{j=2}^{d}\frac{\ell_j(\ell_j+1)}{R_j^2}h_j(\eta)}{
1-\eta\sum_{m=2}^{d}\frac{1}{R_m^2}}.
\end{equation}
The hypergeometric parameters are
\begin{equation}
\begin{aligned}
a&=\frac{1}{2}+i\sqrt{\Xi_s(\eta)-\frac{1}{4}},\\
b&=\frac{1}{2}-i\sqrt{\Xi_s(\eta)-\frac{1}{4}},\\
c&=1+iR_1\omega .
\end{aligned}
\end{equation}
The exponents in the ansatz are chosen as
\begin{equation}
\rho=\frac{c-1}{2},\,\,\,\sigma=\frac{a+b-c}{2}.
\end{equation}
The general solution can be written as
\begin{flalign}\label{Scalar-f}
G(y)= \beta y^{1-c} F(1+a-c, 1+b-c, 2-c; y)+ 
\alpha F(a, b, c; y), 
\end{flalign}
where $F(a,b,c;y)$ denotes the hypergeometric function, and $\alpha$, $\beta$ are integration constants to be fixed by boundary conditions. Substitution of this solution into the radial ansatz gives
\begin{align}\label{defined function-2}
\phi^\omega_{\ell,m}
=(1-y)^{\frac{1}{2}(a+b-c)}
\bigg[
\alpha y^{\frac{1}{2}(c-1)}F(a,b,c;y) 
+\beta y^{-\frac{1}{2}(c-1)}
F(1+a-c,1+b-c,2-c;y)
\bigg].
\end{align}
Near the two horizons, the variable $y$ behaves as
 \begin{equation}\label{yr_boudary}
\left\{
  \begin{array}{ll}
   r_\star \rightarrow -\infty \; \Rightarrow \; y \simeq e^{\frac{2}{R_{1}}\, r_\star} \,, \\
    r_\star \rightarrow +\infty \; \Rightarrow \; (1-y) \simeq   e^{-\frac{2}{R_{1}}\, r_\star} \,.
  \end{array}
\right.
\end{equation}
Since $F(a,b,c;0)=1$, the asymptotic behavior of $\phi^{\omega}_{\ell,m}$ as $r_\star\rightarrow-\infty$ is 
\begin{equation}
 \phi^{\omega}_{\ell,m}|_{r_\star\rightarrow{-\infty}}=\alpha{e^{i\omega{r_\star}}}+\beta{e^{-i\omega{r_\star}}}.
\end{equation}
\begin{figure*}[htb!]
\centering
\includegraphics[width=16.5cm]{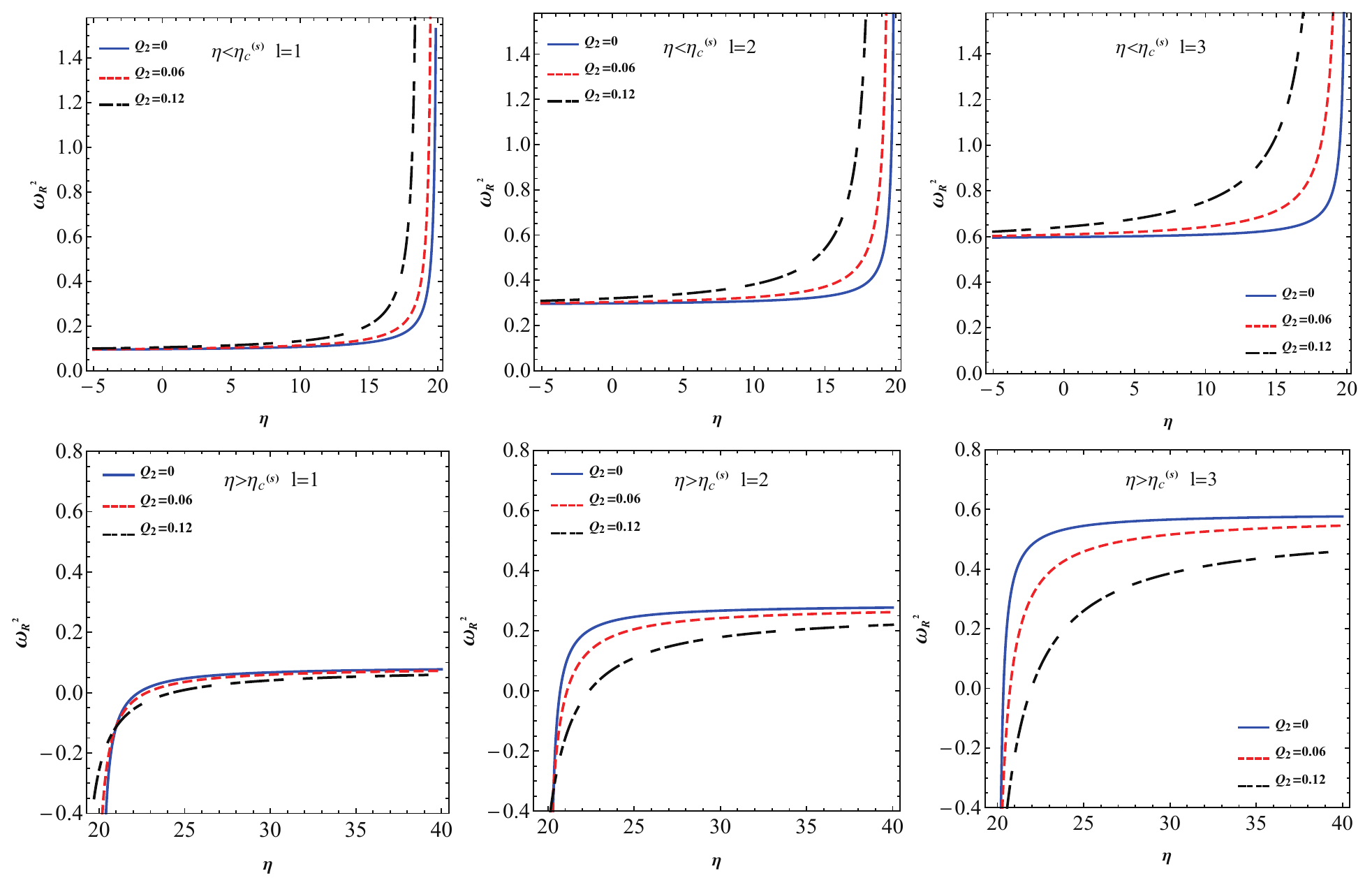}
\caption{Effects of the coupling parameter $\eta$ on the squared real part $\omega_R^2$ of the QNM frequencies of scalar field perturbations in generalized Nariai spacetimes. The upper and lower panels show the branches on the two sides of the corresponding critical value $\eta_c^{(s)}$. Here $D=4$, $Q_1=0.01$, $\Lambda=0.05$ and $\mu=0.01$. }
\label{f1}
\end{figure*}
To impose the boundary conditions near $y=1$, the hypergeometric functions are rewritten in terms of $(1-y)$:
\begin{flalign}
F(a,b,c;y)=
&\frac{\Gamma{(c-a-b)}\Gamma{(c)}}{\Gamma{(c-a)}\Gamma{(c-b)}}F(a,b,a+b-c+1;1-y)+ \nonumber \\
& \frac{\Gamma{(a+b-c)}\Gamma{(c)}}{\Gamma{(a)}\Gamma{(b)}}(1-y)^{(c-a-b)} F(c-a,c-b,c-a-b+1;1-y).
\end{flalign}
The asymptotic behavior of the solution as $r_\star\rightarrow+\infty$ is
\begin{flalign}
\phi^{\omega}_{\ell,m}|_{r_{\star}\rightarrow+\infty}\simeq{C_+e^{i\omega{r_\star}}+C_-e^{-i\omega{r_\star}}},
\end{flalign}
with
\begin{align}
C_+ &=
\frac{\alpha \Gamma(c-a-b)\Gamma(c)}
{\Gamma(c-a)\Gamma(c-b)}
+
\frac{\beta \Gamma(c-a-b)\Gamma(2-c)}
{\Gamma(1-a)\Gamma(1-b)}, \nonumber\\
C_- &=
\frac{\alpha \Gamma(a+b-c)\Gamma(c)}
{\Gamma(a)\Gamma(b)}
+
\frac{\beta \Gamma(a+b-c)\Gamma(2-c)}
{\Gamma(a-c+1)\Gamma(b-c+1)} .
\end{align}
Boundary condition (I) is imposed first. At $r_\star\rightarrow{-\infty}$, the field should describe a purely outgoing wave, which corresponds to the $e^{i\omega{r_\star}}$ mode. Since
\begin{align}
\phi^{\omega}_{\ell,m}|_{r_\star\rightarrow{-\infty}}=\alpha{e^{i\omega{r_\star}}}+\beta{e^{-i\omega{r_\star}}}.
\end{align}
Setting $\beta=0$, the asymptotic form at $r_*\rightarrow+\infty$ reduces to
\begin{align}
\phi^{\omega}_{\ell,m}|_{r_{\star}\rightarrow+\infty}\simeq{e^{i\omega{r_\star}}\alpha{\frac{\Gamma{(c-a-b)}\Gamma{(c)}}{\Gamma{(c-a)}\Gamma{(c-b)}}}}
+e^{-i\omega{r_\star}}\alpha{\frac{\Gamma{(a+b-c)}\Gamma{(c)}}{\Gamma{(a)}\Gamma{(b)}}}.
\end{align}
\begin{figure*}[htb!]
\centering
\includegraphics[width=16.5cm]{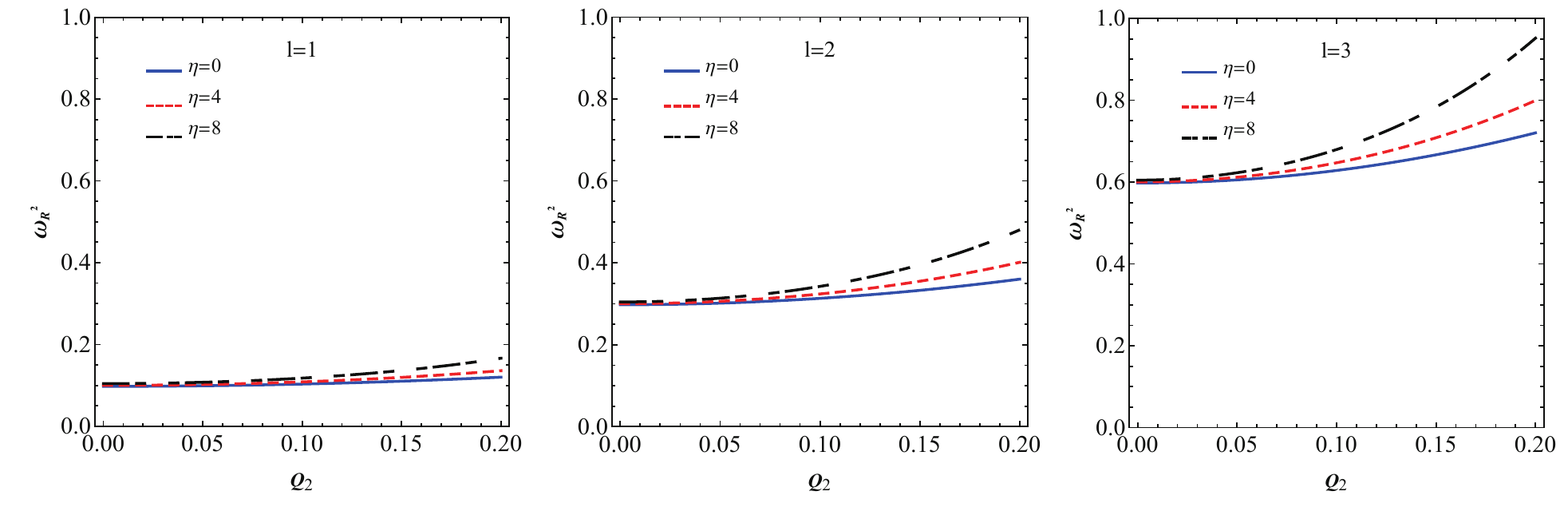}
\caption{Effects of the magnetic charge $Q_2$ on the squared real part $\omega_R^2$ of the QNM frequencies of scalar field perturbations in generalized Nariai spacetimes. Here $D=4$,  $Q_1=0.01$, $\Lambda=0.05$ and $\mu=0.01$.}
\label{f2}
\end{figure*}
At $r_*\to+\infty$, condition (I) requires the coefficient of
$e^{i\omega r_*}$ to vanish. Since $\alpha\neq0$, this is achieved when one of the gamma functions in the denominator diverges, namely
$\Gamma(c-a)\to\infty$ or $\Gamma(c-b)\to\infty$. Under this condition, the corresponding QNM frequencies are
\begin{flalign}
  \omega = \pm\sqrt{\frac{\mu^{2} +\sum\limits_{j=2}^d \frac{\ell_{j}(\ell_j + 1)}{R_j^2}h_j(\eta)}{1-\eta{\sum\limits_{m=2}^{d}\frac{1}{R_{m}^{2}}}} - \frac{1}{4R_1^2}} +\frac{(2n+1)}{2R_1}i,
\end{flalign}
where $n$ is the overtone number. Boundary condition (II) is considered next, for which the mode $\phi^{\omega}_{\ell,m}$ is required to behave as $e^{i\omega{r_\star}}$ at both horizons.A similar analysis shows that this condition requires either $a=-n$ or $b=-n$. Since $a$ and $b$ are independent of $\omega$ in the present parametrization, these conditions do not determine a discrete set of frequencies. Thus, boundary condition (II) does not lead to a well defined QNM spectrum.

Boundary condition (III) is now considered. In this case, the field is required to behave as $e^{-i\omega{r_\star}}$ at both horizons $r_\star\rightarrow\pm\infty$. As $r_\star\rightarrow-\infty$, the general asymptotic form of the radial component is
\begin{equation*}
\phi^{\omega}_{\ell,m}|_{r_\star\rightarrow{-\infty}}=\alpha{e^{i\omega{r_\star}}}+\beta{e^{-i\omega{r_\star}}}.
\end{equation*}
\begin{figure*}[htb!]
\centering
\includegraphics[width=14cm]{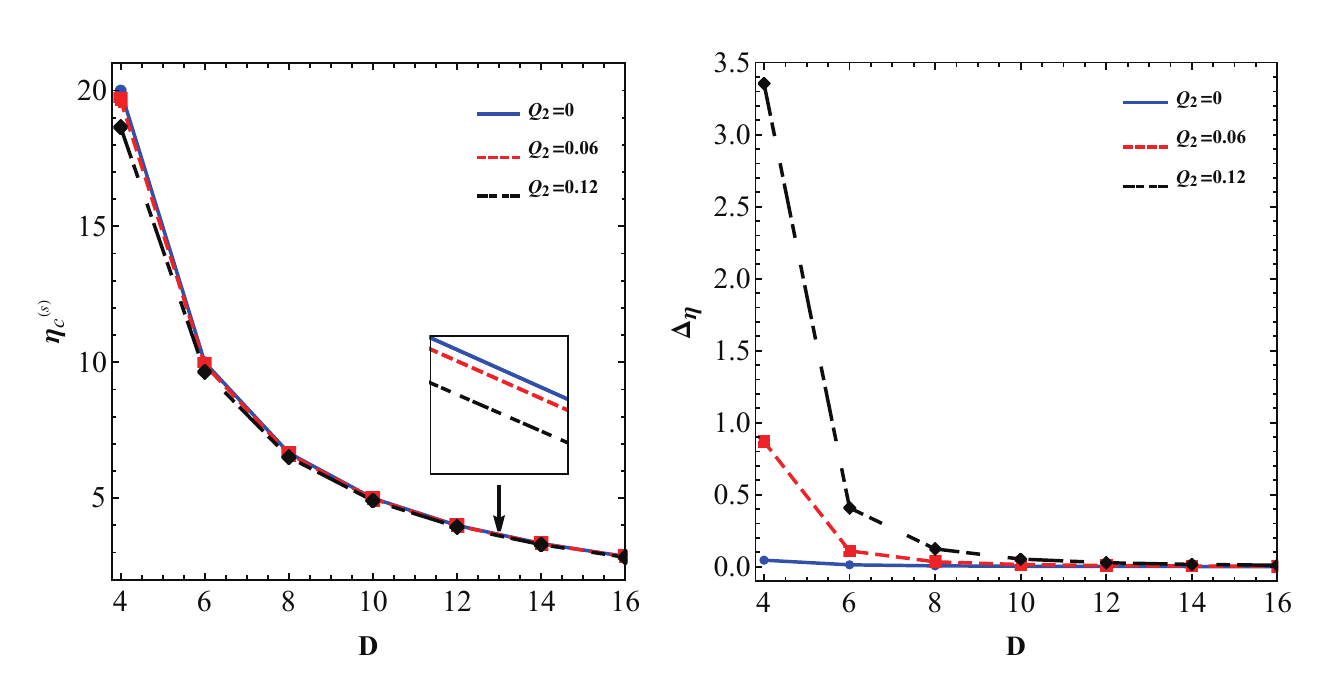}
\caption{Effects of the spacetime dimension $D$ on the critical coupling $\eta_c^{\rm s}$ and the width $\Delta{\eta}$ of scalar field perturbations in generalized Nariai spacetimes. Here $Q_1=0.01$, $\Lambda=0.05$, $\mu=0.01$ and $\ell=1$. For $D>4$, the remaining magnetic charges are fixed as $Q_3=\cdots=Q_d=0.01$.}
\label{f3}
\end{figure*}
Thus, condition (III) imposes $\alpha=0$. The asymptotic behavior at $r_\star=+\infty$ becomes
\begin{align}
\phi^{\omega}_{\ell,m}|_{r_{\star}\rightarrow+\infty}\simeq{e^{i\omega{r_\star}}\beta{\frac{\Gamma{(c-a-b)}\Gamma{(2-c)}}{\Gamma{(1-a)}\Gamma{(1-b)}}}}+e^{-i\omega{r_\star}}\beta{\frac{\Gamma{(a+b-c)}\Gamma{(2-c)}}{\Gamma{(a-c+1)}\Gamma{(b-c+1)}}}.
\end{align}

The conditions $1-a=-n$ or $1-b=-n$ do not fix $\omega$ and therefore do not lead to a discrete QNM spectrum. For boundary condition (IV), an analogous calculation gives the QNM frequencies
\begin{equation}
  \omega = \pm\sqrt{\frac{\mu^{2} +\sum\limits_{j=2}^d \frac{\ell_{j}(\ell_j + 1)}{R_j^2}h_j(\eta)}{1-\eta{\sum\limits_{m=2}^{d}\frac{1}{R_{m}^{2}}}}   - \frac{1}{4R_1^2} } - \frac{(2n+1)}{2R_1}i.
\end{equation}

Analytical expressions for the QNM frequencies are obtained by imposing different boundary conditions. The results show that the stability of the system depends on the choice of boundary condition. For boundary condition (I), the imaginary part of the QNM frequency is positive, indicating exponential growth of perturbations and hence dynamical instability. In contrast, for boundary condition (IV), the imaginary part remains negative, corresponding to damped perturbations and a dynamically stable system. Notably, the squared real part of the frequency, $\omega_R^2$, exhibits an explicit dependence on $\eta$. The effective radial propagation coefficient becomes singular at $\eta_c^{(s)}=\left(\sum_{m=2}^{d}1/R_m^2\right)^{-1}$, which separates the frequency spectrum into two branches. We therefore analyze the behavior of $\omega_R^2$ on both sides of $\eta_c^{(s)}$. In addition, we define $\eta_t$ by $\omega_R^2=0$. The width of this interval is $\Delta\eta=|\eta_t-\eta_c^{(s)}|$. This interval corresponds to the region with $\omega_R^2<0$, where the QNM frequencies become purely imaginary. Thus, $\eta_c^{(s)}$ and $\eta_t$ delimit the oscillatory and purely imaginary QNM regimes.

As shown in Fig. (\ref{f1}), for $\eta<\eta_c^{(s)}$, $\omega_R^2$ increases monotonically with the coupling parameter $\eta$ and diverges as $\eta$ approaches the critical value $\eta_c^{(s)}$. This indicates that the oscillatory part of the mode is amplified as the coupling approaches the critical value. When $\eta_c^{(s)}<\eta<\eta_{t}$, $\omega_R^2$ becomes negative, so that the QNM frequencies become purely imaginary. This implies that the real part of the frequency vanishes, so the modes become nonoscillatory. As $\eta$ increases further, $\omega_R^2$ reappears and grows slowly. We also examine the effect of the charge parameter on the QNM frequencies at fixed coupling, as shown in Fig. (\ref{f2}). The results show that $\omega_R^2$ increases with $Q_2$, indicating that a larger magnetic charge enhances the oscillation frequency of the scalar perturbation.

We further investigate the effect of the spacetime dimension on these characteristic quantities. As shown in Fig. (\ref{f3}), the critical coupling $\eta_c^{(s)}$ decreases rapidly with increasing $D$. Increasing $D$ adds more two sphere factors and enhances the curvature sum entering the radial master equation. As a result, the singular point of the effective radial propagation coefficient moves to smaller coupling values. The left panel shows that the $\eta_c^{(s)}$ curves for different $Q_2$ remain slightly separated in the high dimensional region, although their dependence on $Q_2$ becomes weak. The right panel shows the width $\Delta\eta$ of the purely imaginary QNM regime. This width decreases rapidly as the spacetime dimension increases, indicating that the parameter region where the QNM frequencies are purely imaginary is progressively narrowed in higher dimensional generalized Nariai spacetimes. Moreover, the magnetic charge $Q_2$ enlarges this purely imaginary interval in lower dimensions, especially for $D=4$, but this influence is rapidly suppressed as $D$ increases.

\section{Quasinormal modes of Maxwell Field Perturbations Coupled to the Einstein Tensor}

In this section, we consider Maxwell field perturbations coupled to the Einstein tensor. The Lagrangian density for the Maxwell field is taken as
\begin{equation}\label{Lagrangian-density}
L_M=-\frac{1}{4}\left(F_{\mu\nu}F^{\mu\nu}-4\eta{G^{\mu\rho}}F_{\mu\nu}F_\rho^\nu\right),
\end{equation}
where $F_{\mu\nu}=\partial_\mu\mathcal{A}_\nu-\partial_\nu\mathcal{A}_\mu$ is the Maxwell field strength tensor. Varying the action with respect to $\mathcal{A}_\mu$ yields the modified Maxwell equation in the absence of sources
\begin{align}\label{Maxwell-Einstein}
\nabla_\mu\left(F^{\mu\nu}-2\eta G^{\mu\rho}F_{\rho}^{\ \nu}
+2\eta G^{\nu\rho}F_{\rho}^{\ \mu}\right)=0 .
\end{align}
The last two terms arise from the nonminimal coupling between the Maxwell field and the background Einstein tensor and therefore modify the effective propagation of Maxwell perturbations. Following the standard vector spherical harmonic decomposition for Maxwell perturbations, the axial sector is considered and the vector potential is expanded as
\begin{equation}\label{the four-vector potential}
\begin{split}
\mathcal{A} = e^{-i\omega t} \bigg[  \tilde{A^0} Y_{\ell,m} \, dt + \tilde{A^1} Y_{\ell,m} \, dr_\star 
  - \sum_{j=2}^d \tilde{A^j} \left( \frac{\partial_{\phi_j} Y_{\ell,m}}{\sin\theta_j} d\theta_j - \sin\theta_j \partial_{\theta_j} Y_{\ell,m} \, d\phi_j \right) \bigg],
\end{split}
\end{equation}
here $\tilde{A^0}$, $\tilde A^1$ and $\tilde A^j$ are radial functions of $r_\star$. The pure gradient part has been removed by a gauge transformation. Substituting this ansatz into the modified Maxwell equation gives a coupled radial system, which can be decoupled by introducing
\begin{align}\label{Introduced function}
\Psi_M(r_\star)=\cosh^2(r_\star/R_1)\left[\frac{d}{dr_\star}\tilde{A^0}(r_\star)+i\omega\tilde{A^1}(r_\star)\right].
\end{align}
The time component of Eq.\eqref{Maxwell-Einstein} gives the constraint between $\tilde{A^0}$ and the master variable $\Psi_M$:
\begin{align}\label{relation}
\tilde{A^0}=\left[\frac{\sum_{j=2}^d\frac{q_j(\eta)\ell_j(\ell_j+1)}{R_j^2}}{1+4\eta\sum_{k=2}^d\frac{1}{R_k^2}}\right]^{-1}\frac{d\Psi_M}{dr_\star},
\end{align}
where
\begin{flalign}
 q_j(\eta)=1+2\eta\left(\frac{1}{R_1^2}+\frac{1}{R_j^2}\right)+4\eta\left(\sum_{m=2}^d\frac{1}{R_m^2}-\frac{1}{R_j^2}\right).
 \end{flalign}
Combining the remaining radial equations gives the master equation
\begin{align}\label{maxwell equation}
\left\{\frac{d^2}{dr_\star^2}+\omega^2-\frac{\sum\limits_{j=2}^d\frac{q_j(\eta)\ell_j(\ell_j+1)}{R_j^2}}{\left[1+4\eta\sum\limits_{k=2}^d\frac{1}{R_k^2}\right]\cosh^2(\frac{r_\star}{R_1})}\right\}\Psi_M(r_\star)=0.
\end{align}
Using the same variable $y$ as in the scalar case, the Maxwell master variable is written as
\begin{align}\label{defined function-5}
\Psi_M(r_\star)=y^{\rho}(1-y)^{\sigma}H(y).
\end{align}
Substitution of this ansatz into the radial equation gives the hypergeometric equation
 \begin{align}
  y(1-y)\frac{d^2{H}}{dy^2}+[c-y(a+b+1)]\frac{dH}{dy}-abH=0.
\end{align}
For convenience, we define
\begin{align}
\Xi_M(\eta) \equiv
\displaystyle\sum_{j=2}^{d}
\frac{q_j(\eta)\ell_j(\ell_j+1)R_1^2}{R_j^2}
\left[1+4\eta\displaystyle\sum_{k=2}^{d}\frac{1}{R_k^2}\right]^{-1}.
\end{align}
\begin{figure*}[htb!]
\centering
\includegraphics[width=16.5cm]{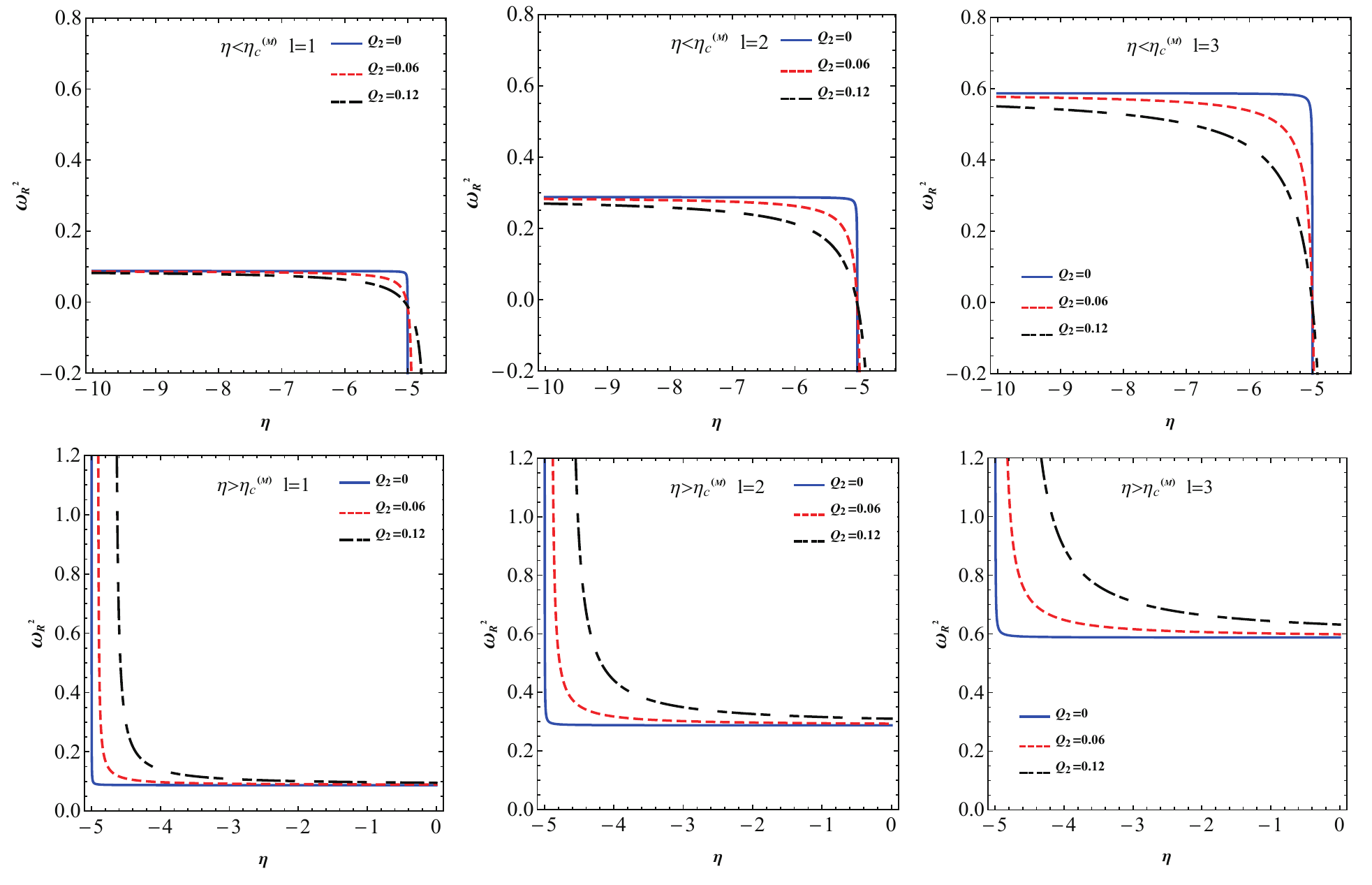}
\caption{Effects of the coupling parameter $\eta$ on the squared real part $\omega_R^2$ of the QNM frequencies of Maxwell field perturbation in generalized Nariai spacetimes. The upper and lower panels show the branches on the two sides of the corresponding critical value $\eta_c^{\scriptscriptstyle (M)}$. Here $Q_1=0.01$ and $\Lambda=0.05$.}
\label{f4}
\end{figure*}
\begin{figure*}[htb!]
\centering
\includegraphics[width=16.5cm]{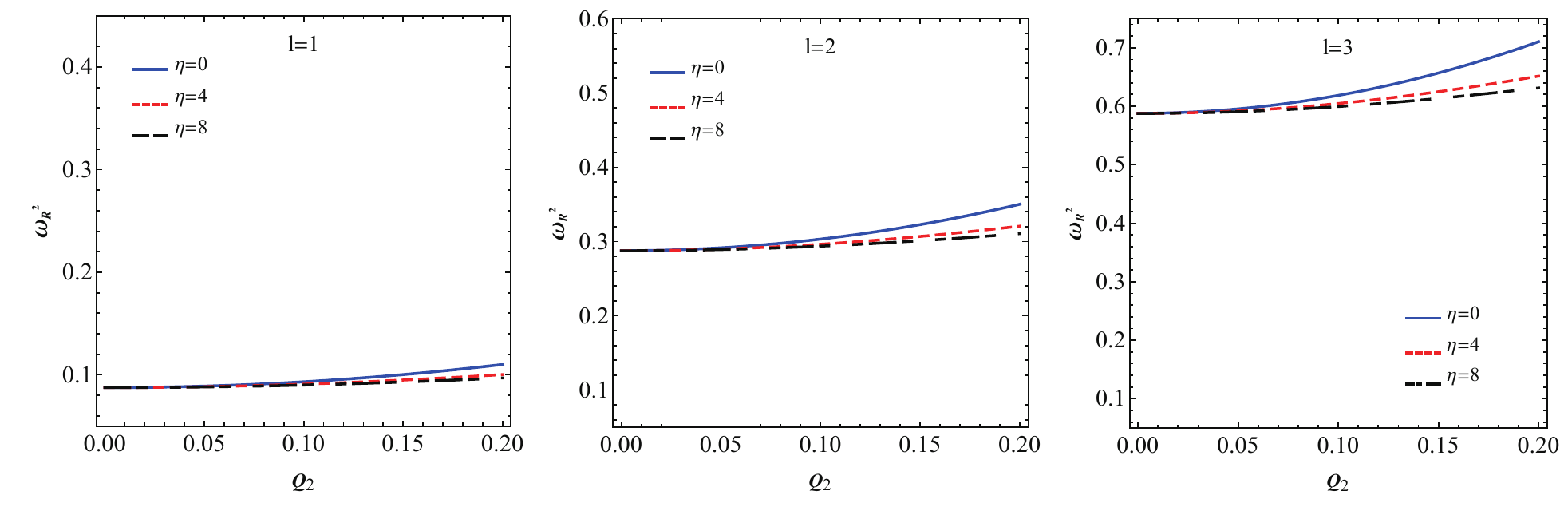}
\caption{Effects of the magnetic charge $Q_2$ on the squared real part $\omega_R^2$ of the QNM frequencies of Maxwell field perturbation in generalized Nariai spacetimes. Here $Q_1=0.01$ and $\Lambda=0.05$.}
\label{f5}
\end{figure*}
The corresponding hypergeometric parameters are
\begin{equation}
\begin{aligned}
a&=\frac{1}{2}+i\sqrt{\Xi_M(\eta)-\frac{1}{4}},\\
b&=\frac{1}{2}-i\sqrt{\Xi_M(\eta)-\frac{1}{4}},\\
c&=1+iR_1\omega .
\end{aligned}
\end{equation}
\begin{figure*}[htb!]
\centering
\includegraphics[width=14cm]{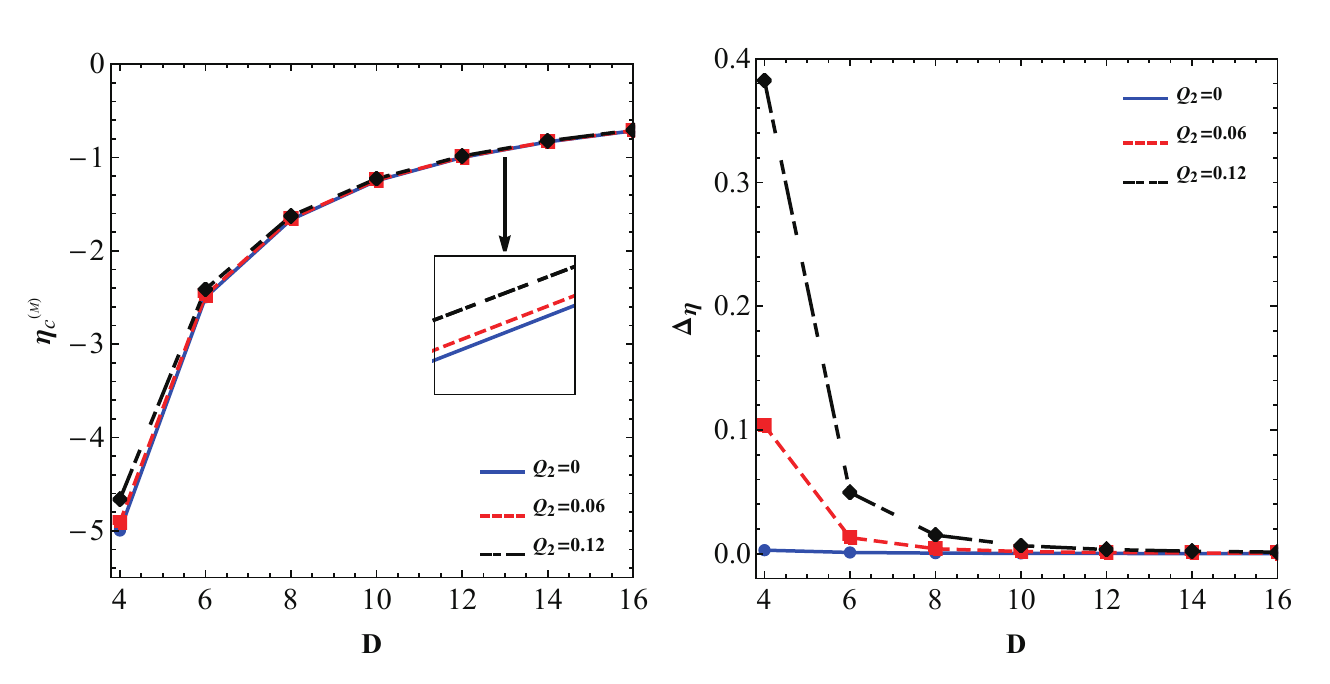}
\caption{Effects of the spacetime dimension $D$ on the critical coupling $\eta_c^{\scriptscriptstyle (M)}$ and the width $\Delta\eta$ of Maxwell field perturbations in generalized Nariai spacetimes. Here $Q_1=0.01$, $\Lambda=0.05$ and $\ell=1$. For $D>4$, the remaining magnetic charges are fixed as $Q_3=\cdots=Q_d=0.01$.}
\label{f6}
\end{figure*}  
The exponents in the ansatz are
\begin{equation}
\rho=\frac{c-1}{2},\,\,\,\sigma=\frac{a+b-c}{2}.
\end{equation}
The general solution for $H(y)$ is
\begin{flalign}\label{H}
\begin{split}
H(y)=\beta{y^{(1-c)}}F(1+a-c,1+b-c,2-c;y)+\alpha{F(a,b,c;y)}.
\end{split}
 \end{flalign}
Substitution of the general solution for $H(y)$ into the ansatz for $\Psi_M$ gives 
\begin{align}
\Psi_M=(1-y)^{\frac{1}{2}(a+b-c)}
\bigg[
\alpha y^{\frac{1}{2}(c-1)}F(a,b,c;y)
+\beta y^{-\frac{1}{2}(c-1)}
F(1+a-c,1+b-c,2-c;y)
\bigg].
\end{align}
Imposing the same set of boundary conditions on the Maxwell master variable gives the corresponding QNM frequencies. For the admissible boundary conditions, the pole conditions are $c-a=-n$ or $c-b=-n$, and hence
\begin{align}\label{the quasinormal frequencies-M}
\omega = \pm\sqrt{\frac{\sum\limits_{j=2}^d\frac{q_j(\eta)\ell_j(\ell_j+1)}{R_j^2}}{1+4\eta\left(\sum\limits_{k=2}^d\frac{1}{R_k^2}\right)} - \frac{1}{4R_1^2} } \pm\frac{i}{2R_1} (2n+1).
\end{align}
As in the scalar case, the two signs of the explicit imaginary term correspond to the two admissible boundary conditions. From Eq. \eqref{the quasinormal frequencies-M}, the coupling parameter $\eta$ modifies the square root term of the Maxwell QNM frequencies, while the explicit imaginary term is independent of $\eta$. The effective radial coefficient becomes singular at $\eta_c^{\scriptscriptstyle (M)}=-\left(4\sum_{k=2}^{d}\frac{1}{R_k^2}\right)^{-1}$, which separates the frequency spectrum into two branches. The behavior of $\omega_R^2$ is then analyzed on the two sides of this critical value. As shown in Fig.(\ref{f4}), for $\eta<\eta_c^{\scriptscriptstyle (M)}$, $\omega_R^2$ decreases monotonically as $\eta$ increases and approaches zero at $\eta_t$. In the interval $\eta_t<\eta<\eta_c^{\scriptscriptstyle (M)}$, $\omega_R^2$ becomes negative, and the QNM frequencies become purely imaginary. For $\eta>\eta_c^{\scriptscriptstyle (M)}$, $\omega_R^2$ becomes positive again and decreases slowly as $\eta$ increases, as shown in the lower panels of Fig. \ref{f4}. Fig.(\ref{f5}) shows the effect of the magnetic charge $Q_2$ on $\omega_R^2$ at fixed values of the coupling parameter. The results indicate that $\omega_R^2$ generally increases with $Q_2$, implying that a larger magnetic charge enhances the oscillatory response of the Maxwell perturbations.

We further examine how the characteristic quantities of Maxwell perturbations depend on the spacetime dimension. As shown in Fig.(\ref{f6}), the critical value $\eta_c^{\scriptscriptstyle (M)}$ is negative in the Maxwell case, in contrast to the scalar field case. The left panel shows that $\eta_c^{\scriptscriptstyle (M)}$ increases toward zero as the spacetime dimension increases. This indicates that the singular point of the effective radial propagation coefficient is shifted to weaker negative coupling in higher dimensions. The right panel shows the width $\Delta\eta$ of the purely imaginary QNM regime of Maxwell perturbations. In contrast to the scalar case, this purely imaginary interval lies in the region $\eta_t<\eta<\eta_c^{\scriptscriptstyle (M)}$. The rapid decrease of $\Delta\eta$ with increasing $D$ indicates that the purely imaginary Maxwell QNM regime is substantially narrowed in higher dimensional generalized Nariai spacetimes. Moreover, the effect of the magnetic charge $Q_2$ is most visible in lower dimensions and becomes much weaker as $D$ increases.

\section{conclusions}

In this work, we investigated the QNMs of scalar and Maxwell field perturbations coupled to the Einstein tensor in generalized Nariai spacetimes. By imposing different boundary conditions at the two Killing horizons, analytical expressions for the corresponding QNM frequencies were obtained. The results show that the stability of the perturbations depends on the choice of boundary condition. Under boundary condition (I), the imaginary part of the frequency is positive, so that the perturbation grows exponentially in time. By contrast, under boundary condition (IV), the imaginary part is negative and the perturbation is damped. Thus, only the modes obtained under appropriate boundary conditions correspond to stable QNMs.

For scalar field perturbations, the coupling parameter $\eta$ affects the squared real part $\omega_R^2$ of the frequency. The effective radial propagation coefficient reaches a singular point at $\eta_c^{(s)}=\left(\sum_{m=2}^{d}\frac{1}{R_m^2}\right)^{-1}$, which separates the spectrum into two branches. In the branch $\eta<\eta_c^{(s)}$, $\omega_R^2$ increases with $\eta$ and diverges as $\eta$ approaches $\eta_c^{(s)}$. In the branch $\eta>\eta_c^{(s)}$, there exists an interval $\eta_c^{(s)}<\eta<\eta_t$, in which $\omega_R^2<0$ and the QNM frequencies become purely imaginary. As $\eta$ increases further, $\omega_R^2$ becomes positive again and the oscillatory behavior is restored. For Maxwell field perturbations, the Einstein tensor coupling also modifies $\omega_R^2$, but the critical structure differs from that of the scalar field. The corresponding effective radial propagation coefficient reaches its singular point at $\eta_c^{\scriptscriptstyle (M)}=-\left(4\sum_{k=2}^{d}\frac{1}{R_k^2}\right)^{-1}$, so the corresponding critical coupling is negative. In this case, the purely imaginary frequency interval lies in $\eta_t<\eta<\eta_c^{\scriptscriptstyle (M)}$. A comparison of the two fields shows that the coupling affects their oscillatory parts in opposite ways: away from the purely imaginary regime, $\omega_R^2$ generally increases with $\eta$ for the scalar field, whereas it decreases with $\eta$ for the Maxwell perturbations. 

We further analyzed the effects of the magnetic charge and the spacetime dimension on these critical behaviors. For both types of perturbations, the magnetic charge $Q_2$ generally enhances $\omega_R^2$, corresponding to a stronger oscillatory response to external perturbations. The magnetic charge also changes the width of the purely imaginary QNM regime more significantly in lower dimensions, whereas its influence is rapidly weakened in higher dimensions. As the spacetime dimension $D$ increases, the width $\Delta\eta$ of the corresponding purely imaginary QNM regime decreases rapidly for both scalar and Maxwell perturbations. This indicates that higher dimensional generalized Nariai spacetimes shrink the parameter region in which the QNM frequencies are purely imaginary, thereby suppressing the coupling induced nonoscillatory regime.

Overall, these results indicate that nonminimal coupling to the Einstein tensor can modify both the effective propagation and the QNM frequencies of perturbing fields in generalized Nariai spacetimes. Scalar and Maxwell field perturbations share similar critical behaviors induced by the coupling, but they also exhibit important differences determined by the spin of the field and the specific coupling structure. These results help clarify the combined effects of curvature coupling, background electromagnetic parameters, and spacetime dimension on QNM frequencies and stability in higher dimensional near horizon geometries.


\section*{Acknowledgments}
This work was supported by the National Natural Science Foundation of China (Grant Nos.12405053, 12275079, 12035005,12375045), the National Key Research and Development Program of China (Grant No. 2020YFC2201400), and the innovative research group of Hunan Province (Grant No. 2024JJ1006).


\begin{thebibliography}{99}
\bibitem{LG1} B. P. Abbott et al. (LIGO Scientific, Virgo), \textit{Observation of Gravitational Waves from a Binary Black Hole Merger}, Phys. Rev. Lett. {\bf116}, 061102 (2016).
\bibitem{LG2} B. P. Abbott et al. (LIGO Scientific, Virgo), \textit{GW170817: Observation of Gravitational Waves from a Binary
Neutron Star Inspiral}, Phys. Rev. Lett. {\bf119}, 161101 (2017).
\bibitem{LG3} R. Abbott et al. (KAGRA, VIRGO, LIGO Scientific), \textit{GWTC-3: Compact Binary Coalescences Observed by LIGO
and Virgo during the Second Part of the Third Observing Run}, Phys. Rev. X {\bf13}, 041039 (2023)


\bibitem{QM1}  A. Bonanno, R. A. Konoplya, G. Oglialoro, A. Spina, \textit{Regular Black Holes from Proper-Time flow in Quantum Gravity and their Quasinormal modes, Shadow and Hawking radiation}. J. Cosmo. Astro. Phys. {\bf12}, 042 (2025).
\bibitem{QM2} M. F. S. Alves, B. P. P$\hat{o}$nquio, L. G. Medeiros ,\textit{Critical masses and numerical computation of massive scalar quasinormal modes in Schwarzschild black holes}, Phys. Rev. D {\bf112}, 124007 (2025).
\bibitem{QM3} F. Ahmed, S. N. Gashti, A. Bouzenada, B. Pourhassan, \textit{Schwarzschild-AdS Black Holes with Cloud of Strings and Quintessence: Geodesics, Thermodynamic Topology, and Quasinormal Modes}, Nucl. Phys. B, 117260 (2025).
\bibitem{QM4} W. Deng, W. Liu, F. Long, K. Xiao, J. Jing, \textit{Quasinormal Modes of a Massive Scalar Field in Slowly Rotating Einstein-Bumblebee Black Holes}, J. Cosmo. Astro. Phys. {\bf11}, 028 (2025).
\bibitem{QM5} Q. Tan, S. Long, W. Deng, and J. Jing, \textit{Quasinormal modes and echoes of a double braneworld}, J. High Energ. Phys. {\bf02}, 055 (2025).
\bibitem{QM6} C. Chen, J. Jing, Z. Cao, M. Wang, \textit{Complete quasinormal modes of Type-D black holes}, Phys. Rev. D {\bf112}, 103036 (2025).
\bibitem{QM7} F. Naderi, A. Rezaei-Aghdam, \textit{Quasinormal modes of three-dimensional black holes in string theory, conformal gravity, and Hu-Sawicki  theory via the Heun function}, Eur. Phys. J. C  {\bf84}, 1283 (2024).
\bibitem{QM8} L. Zhu, G. Fu, S. Li, D. Zhang, J. Wu, \textit{Quasinormal modes of a charged loop quantum black hole}, Phys.Rev.D 111  {\bf10}, 104008 (2025).
\bibitem{QM9} N. Herceg, T. Juri$\Acute{c}$, A. N. Kumara, A. Samsarov, I. Smoli$\Acute{c}$, \textit{Noncommutative quasinormal modes of Schwarzschild black hole}, J. High Energ. Phys. {\bf2025}, 83 (2025).
\bibitem{QM10} Z. Yang, Y. Lei, X. Kuang, B. Wang, \textit{Gravitational odd-parity perturbation of a Horndeski hairy black hole: quasinormal mode and parameter constraint}, Eur. Phys. J. C {\bf85}, 100 (2025)
\bibitem{QM11} T. Chen, R. Cai, B. Hu, \textit{Quasinormal modes of gravitational perturbation for uniformly accelerated black holes}, Phys. Rev. D {\bf109}, 084049 (2024).
\bibitem{QM12} K. Lin, \textit{Quasinormal modes and echo effect of cylindrical anti-de Sitter black hole spacetime with a thin shell}, Phys. Rev. D {\bf107}, 124002 (2023).

\bibitem{xll} S. Mukhi, \textit{String theory: A perspective over the last 25 years}, Classical Quant. Grav. {\bf28}, 153001 (2011).

\bibitem{msj1} N. Arkani, S. Dimopoulos, and G. R. Dvali, \textit{The Hierarchy problem and new dimensions at a millimeter}. Phys. Lett. B, {\bf429}, 263 (1998).
\bibitem{msj2} I. Antoniadis, N. Arkani-Hamed, S. Dimopoulos, and G. Dvali, \textit{New dimensions at a millimeter to a Fermi and superstrings at a TeV}. Phys. Lett. B, {\bf436}, 257 (1998).
\bibitem{msj3} L. Randall and R. Sundrum. \textit{An Alternative to compactification}. Phys. Rev. Lett., {\bf83}, 4690 (1998).

\bibitem{ac1} Maldacena, \textit{The large-N limit of superconformal field theories and supergravity}, Int. J. Theor. Phys. {\bf38}, 1113 (1999);
\bibitem{ac2} V. E. Hubeny, \textit{The AdS/CFT correspondence}, Classical Quant. Grav. {\bf32}, 124010 (2015).


\bibitem{Nariai1} H. Nariai, \textit{On a new cosmological solution of Einstein’s field equations of gravitation}, Sci. Rep. Tohoku Univ. {\bf35}, 62 (1951).
\bibitem{Nariai2} H. Kodama ,A. Ishibashi, \textit{Master equations for perturbations of generalized static black holes with charge in higher dimensions}, Prog. Theor. Phys. {\bf111}, 29 (2004).
\bibitem{Nariai3} Carlos Batista1, \textit{Generalized charged Nariai solutions in arbitrary even dimensions with multiple magnetic charges}, Gen. Relativ. Gravit. {\bf48}, 160 (2016).
\bibitem{Nariai4} J. Ven$\hat{a}$ncio and C. Batista, \textit{Quasinormal modes in generalized Nariai spacetimes}, Phys. Rev. D {\bf97}, 105025 (2018).

\bibitem{oh1} M. Karimabadi, D. M. Yekta and S. A. Alavi, \textit{Scalar field perturbations in Non-commutative Schwarzschild spacetime: Comparative analysis and Upper bound on non-commutativity}.arXiv:2508.13820.
\bibitem{oh2} S. Fan, W. Guo and C. Wu. Grey-body Factors and Absorption Cross Sections of Non-Commutative Black Holes under Einstein-Coupled Scalar Fields. Class. Quantum Grav. {\bf43}, 105024 (2026).
\bibitem{oh3} N. Chatzifotis, C. Vlachos, K. Destounis and E. Papantonopoulos, \textit{Stability of black holes with non-minimally coupled scalar hair to the Einstein tensor}, Gen. Relativ. Gravit. {\bf54}, 49 (2022).
\bibitem{oh4} D. Wang, M. Sun, Q. Pan, J. Jing, \textit{Backreacting holographic superconductors from the coupling of a scalar field to the Einstein tensor}, Phys. Lett. B {\bf785}, 362 (2018).
\bibitem{oh5} M. Sun, D. Wang, Q. Pan, J. Jing, \textit{Generalized superconductors from the coupling of a scalar field to the Einstein tensor and their refractive index in massive gravity}, Eur. Phys. J. C {\bf79}, 145 (2019).
\bibitem{oh6} X. Kuang, E. Papantonopoulos, \textit{Building a Holographic Superconductor with a Scalar Field Coupled Kinematically to Einstein Tensor}, J. High Energy Phys. {\bf08}, 161 (2016).
\bibitem{oh7} S. Chen and J. Jing, \textit{Dynamical evolution of a scalar field coupling to Einstein's tensor in the Reissner-Nordström black hole spacetime}, Phys. Rev. D {\bf82}, 084006 (2010).
\bibitem{oh8} R. Konoplya, Z. Stuchl{\'i}k, and A. Zhidenko, \textit{Massive nonminimally coupled scalar field in Reissner-Nordström spacetime: Long-lived quasinormal modes and instability}, Phys. Rev. D {\bf98}, 104033 (2018).
\bibitem{oh9} E. Abdalla, B. Cuadros-Melgar, R. Fontana, J. de Oliveira, E. Papantonopoulos, and A. Pavan, \textit{Instability of Reissner-Nordström-AdS black hole under perturbations of a scalar field coupled to Einstein tensor}, Phys. Rev. D 99, 104065 (2019).
\bibitem{oh10} S. Chen and J. Jing. \textit{Dynamical evolution of the electromagnetic perturbation with Weyl corrections}, Phys. Rev. D 88, 064058 (2013).
\bibitem{oh11} R. Kase, M. Minamitsuji, S. Tsujikawa and Y. Zhang, \textit{Black hole perturbations in vector-tensor theories: The odd-mode analysis}, J. Cosmo. Astro. Phys. {\bf02} 048 (2018).
\bibitem{oh12} C. Chen, A. D. Felice and S. Tsujikawa. Linear stability of vector Horndeski black holes. J. Cosmo. Astro. Phys. {\bf07} 022 (2024).
\bibitem{oh13} S. Chen and J. Jing, \textit{Dynamical evolution of a vector field perturbation coupled to the Einstein tensor}, Phys. Rev. D {\bf90}, 124059 (2014).

\bibitem{js1} R. Dutt, A. Khare, and U. P. Sukhatme, \textit{Supersymmetry, shape invariance, and exactly solvable potentials}, Am. J.
Phys. {\bf56}, 163 (1988).

\end{thebibliography}
\end{document}